\documentclass[12pt,preprint,natbib]{emulateapj}

\newcommand{\be} {\begin{equation}}

\def\sgr{SGR\,0418+5729}

\def\swt {Swift\,J1955}

\newcommand{\CXO}{{\it Chandra}\,}

\newcommand{\Swift}{{\em Swift}}
\newcommand{\bc}{\begin{center}}
\newcommand{\ec}{\end{center}}
\def\ltsima{$\; \buildrel < \over \sim \;$}
\def\lsim{\lower.5ex\hbox{\ltsima}}
\def\loe{\lower.5ex\hbox{\ltsima}}
\def\gtsima{$\; \buildrel > \over \sim \;$}
\def\gsim{\lower.5ex\hbox{\gtsima}}
\def\goe{\lower.5ex\hbox{\gtsima}}

\def\ltsima{$\; \buildrel < \over \sim \;$}
\def\lsim{\lower.5ex\hbox{\ltsima}}
\def\loe{\lower.5ex\hbox{\ltsima}}
\def\gtsima{$\; \buildrel > \over \sim \;$}
\def\gsim{\lower.5ex\hbox{\gtsima}}
\def\goe{\lower.5ex\hbox{\gtsima}}

\def\ergs {erg\,s$^{-1}$}
\def\ergscm2 {erg\,s$^{-1}$cm$^{-2}$}

\def\cm2 {cm$^{-2}$}
\def\arcsec{$^{\prime\prime}$}

\shortauthors{REA ET AL.}
\shorttitle{The quiescence of Swift\,J195509.6+261406 (GRB\,070610)}

\begin{document}

\title{The X-ray quiescence of Swift\,J195509.6+261406 (GRB\,070610): \\ an optical bursting X-ray binary?}


\author{N. Rea\altaffilmark{1}, P.~G. Jonker\altaffilmark{2,3,4}, G. Nelemans\altaffilmark{4}, J.~A. Pons\altaffilmark{5}, M. M. Kasliwal\altaffilmark{6}, S. R. Kulkarni\altaffilmark{6}, R. Wijnands\altaffilmark{7}}

\altaffiltext{1}{Institut de Ciencies de l'Espai (CSIC--IEEC), Campus UAB, Facultat de Ciencies, Torre C5-parell, 2a planta, 08193, Bellaterra (Barcelona), Spain; Email: rea@ieec.uab.es}
\altaffiltext{2} {SRON-Netherlands Institute for Space Research, Sorbonnelaan 2, 3584 CA Utrecht, the Netherlands} 
\altaffiltext{3} {Harvard-Smithsonian Center for Astrophysics, 60 Garden Street, Cambridge, MA 02138, USA} 
\altaffiltext{4} {Radboud University Nijmegen, Department of Astrophysics, IMAPP, PO Box 9010, 6500 GL, Nijmegen, The Netherlands}
\altaffiltext{5} {Departament de Fisica Aplicada, Universitat d'Alacant, Ap. Correus 99, 03080 Alacant, Spain}
\altaffiltext{6} {California Institute of Technology, Department of Astronomy, Mail Stop 105-24,  Pasadena, CA 91125, USA} 
\altaffiltext{7} {University of Amsterdam, Astronomical Institute ``Anton Pannekoek'',  Postbus 94249, 1090 GE, Amsterdam, The Netherlands} 

\begin{abstract}

We report on a $\sim$63\,ks \CXO\, observation of the X-ray transient
Swift\,J195509.6+261406\, discovered as the afterglow of what
was first believed to be a long duration Gamma-Ray Burst (GRB\,070610). The outburst of this source was 
characterized by unique optical flares on timescales of second or less, morphologically similar to the short X-ray bursts usually observed from magnetars. Our \CXO\, observation was performed $\sim$2 years after the discovery of
the optical and X-ray flaring activity of this source, catching it in
its quiescent state. We derive stringent upper limits on the
quiescent emission of Swift\,J195509.6+261406 which argues
against the possibility of this object being a typical magnetar. Our limits show
that the most viable interpretation on the nature of this peculiar
bursting source, is a binary system hosting a black hole or a neutron
star with a low mass companion star ($< 0.12 M_{\odot}$), and with an
orbital period smaller than a few hours.

\end{abstract}


\section{INTRODUCTION}
\label{intro}

On 2007 June 10 the \Swift\, Burst Alert Telescope (BAT) triggered on
GRB\,070610, a typical long-duration GRB (see Gehrels et al.~2007 for
a recent review), with a $\sim$4.6\,s high-energy prompt emission
(Pagani et al.~2007; Tueller et al.~2007). Follow-up soft X-ray
observations with the \Swift\, X-ray Telescope (XRT) started soon after the event, discovering
only one variable X-ray source within the BAT error circle:
namely Swift\,J195509.6+261406\ (Kasliwal et
al.~2008; \swt\, hereafter). This transient X-ray source was
very different from what expected for the X--ray afterglow of a long
GRB: it was decreasing in flux rather slowly, and it showed 
a strong X-ray flaring activity.

The source became undetectable by \Swift-XRT on 2007 June 29, ranging
from a 0.5--10\,keV flux of $\sim10^{-9}$ to
$< 10^{-12}$\ergscm2 \, in 19 days. While in outburst, \swt\, had
an X-ray spectrum that could be described by a rather hard power-law
corrected for the photoelectric absorption ($N_{\rm H}
=7\times10^{21}$\cm2 \, and $\Gamma$=1.7). Due to spatial and
temporal coincidence (it was the only transient source in the BAT
error circle), GRB\,070610 and the X--ray transient \swt\, have been
associated with high probability (Kasliwal et al.~2008).

The most interesting and peculiar features of this transient
source came from optical and infrared observations. Many telescopes,
triggered by the GRB-like event, promptly observed the position of
\swt\, during the outburst. A highly variable optical and infrared counterpart was observed, showing large
flares for about 11 days after the GRB-like event, when it 
went back to quiescence. These large flares were characterized by a
very short timescale: during the largest flare the source increased
its optical flux by more than a factor of 200 in less than 4\,s. Furthermore, a broad quasi periodic
oscillation was observed in the optical band at $\sim$0.16\,Hz
(Stefanescu et al. 2008).

The source distance was constrained by several different methods to be within
3.7--10\,kpc (mainly red clump study, and detailed measurements of the absorption column in the {\em mm}
waveband; Castro-Tirado et al. 2008). Furthermore, the stringent optical and IR limits derived
in the quiescent level (H$>$23; R$>$26.0 and $i^{\prime}>$24.5;
Kasliwal et al. 2008; Castro-Tirado et al. 2008) constrain the type
of any companion star to either a main-sequence star with spectral
type later than M5V (which means a mass $<$0.12~M$_{\odot}$), or to a
semi-degenerate hydrogen poor star (Castro-Tirado et al. 2008).

The large variations of its optical and infrared counterpart during
the decay to quiescence, its distance and Galactic nature, set 
this transient apart from the typical optical afterglows of  long-duration GRBs (see Liang et al.~2007 for a recent review).

The resemblance of the optical bursts of \swt\,
with the short X-ray bursts from magnetars (see Mereghetti 2008 for a recent
review) led to the idea of a new kind of X-ray and optical transient
event in a Galactic magnetar (Castro-Tirado et al. 2008; Stefanescu et
al. 2008). On the other hand, its X-ray flaring activity was also proposed to
resemble the emission of the fast X-ray nova V4641 Sgr (Markwardt et al. 2008; Kasliwal et
al. 2008), an unusual 9~M$_{\odot}$ black hole in orbit with a 5--8~M$_{\odot}$
B9 III companion star (in't Zand et al. 2000; Orosz et al. 2001)

In this Letter we present the results of a $\sim$63\,ks \CXO\,
observation of \swt\, (see \S\,\ref{obs}) aimed at unveiling its X-ray
properties during quiescence (see \S\,\ref{results}), and compare them
with the current quiescent levels of the magnetar and X-ray binary
populations (see \S\,\ref{discussion}).

\section{OBSERVATION AND DATA ANALYSIS}
\label{obs}

The \CXO\, X-ray Observatory observed \swt\, for $\sim$63\,ks with the
Advanced CCD Imaging Spectrometer (ACIS) instrument (ObsID\,10042)
from 2009 August 03 16:09:57 to August 04 09:55:59 (Terrestrial Time) in {\tt VERY
  FAINT (VF)} timed exposure imaging mode.  The source was positioned
on the back-illuminated ACIS-S3 CCD at the nominal target position (RA: 19 55 09.653, Dec: +26 14 05.84 $\pm$0\farcs27; J2000), and we used a sub-array of 1/8
leading to a time resolution of 1.14\,s. Standard processing of the
data was performed by the \CXO\, X-ray Center to Level 1 and Level 2
(processing software DS 8.0). The data were reprocessed using the CIAO
software (version 4.1.2). We used the latest ACIS gain map, and
applied the time-dependent gain and charge transfer inefficiency
corrections. The data were then filtered for bad event grades and only
good time intervals were used. No high background events were
detected, resulting in a final on-time exposure of 62.732\,ks.

\section{RESULTS}
\label{results}

We did not detect any X-ray source at the position of the optical
counterpart to \swt\, (Kasliwal et al. 2008; Castro-Tirado et
al. 2008).
In particular we detected no 0.3--10\,keV photon within a 1\arcsec\,
circle centred on the optical position. We took a 95\% upper limit of
3 photons (Gehrels 1986) and inferred a
$4.78\times10^{-5}$~counts~s$^{-1}$ upper limit on the X-ray quiescent
count-rate of \swt . 

Using {\tt
  PIMMS}\footnote{http://heasarc.gsfc.nasa.gov/Tools/w3pimms.html} we
estimated the 95\% upper limit on the source flux assuming: a) an absorbed power-law spectrum similar to the outburst  spectral energy distribution ($N_{\rm H}
=7\times10^{21}$\cm2 \, and $\Gamma$=1.7; Kasliwal et al. 2008), and  b) a quiescent thermal spectrum with the same $N_{\rm H}
=7\times10^{21}$\cm2 \, and kT=0.3\,keV, typical of a magnetar
in quiescence (see i.e., Muno et al. 2008; Bernardini et al. 2009). We
derived a 95\% upper limit on the observed 0.3--10\,keV absorbed (unabsorbed)
flux of $6.2 (9.6)\times10^{-16}$\ergscm2 \, and $2.1
(6.6)\times10^{-16}$\ergscm2 , under the power-law and blackbody
spectral assumption, respectively. Note that these spectral decompositions, and the derived flux limits, comprise also the typical values for quiescent X-ray binaries.


\section{DISCUSSION}
\label{discussion}

We present in this Letter a deep X-ray observation of \swt\, during the quiescent state. Before entering in a detailed discussion of our X-ray limit on the quiescent emission of this optical bursting transient,  we first
need to discuss the current estimates on the source
distance. 

Castro-Tirado et al. (2008) studied in detail the distance
issue, first taking $^{12}$CO and H {\tt I} spectra, and then deriving
the extinction versus distance distribution in the \swt\, line of sight
based on the red clump method (see also Lopez-Corredoira et al. 2002;
Durant \& van Kerkwijk 2006). From the mm and cm spectra,
they derived a Galactic column density in the direction of this source
of $N_{\rm H} = N_{\rm H I} + 2N_{\rm H II} = 14.1 \pm
2.0\times 10^{21}$\cm2 ,half of which is accounted for by a molecular cloud at 3.7\,kpc (considered as a lower limit on the
distance of \swt). Comparing the Galactic $N_{\rm H}$ with the one
derived from fitting the X-ray spectrum during outburst ($N_{\rm
  H}=7.2^{+3}_{-2} \times 10^{21}$\cm2 ; Kasliwal et
al. 2008)\footnote{Note that this is derived from an absorbed
  power-law fit (Kasliwal et al. 2008), hence it might be an
  overestimate on the $N_{\rm H}$ of the source.}, \swt\, is expected
to be located at $\sim$4--5\,kpc. Similar results have been derived
from the red-clump method, from which a distance of $\sim$4\,kpc could
be inferred (Castro-Tirado et al. 2008). Unfortunately the upper limit
on the source distance is not very well constrained, although all
methods used would place the source $\sim$4--5\,kpc. The main problem
to assess a distance error bar is that the Galactic plane in the
direction of \swt\, extends only until
$\sim$5\,kpc, behind which there is the bulge, with its intrinsically
lower column density. This makes it extremely hard to define an upper limit to the 
distance, since in the bulge a column density versus distance relation
is not well defined. Hereafter, we will discuss our results
assuming a distance range of 3.7--10\,kpc, considering the farthest limit in the
Milky Way as a distance upper limit.

The 95\% upper limit on the quiescent X-ray luminosity of \swt\, 
is $2.8\times10^{30}$d$_{\rm 5kpc}^2$\ergs \, or $1.9\times10^{30}$d$_{\rm 5kpc}^2$\ergs , for the
power-law or blackbody models, respectively (see also
\S\ref{results}). Considering the whole 3.7--10\,kpc range
(see above), the derived quiescent X-ray luminosity range is between
1.5--11.4$\times10^{30}$\ergs\, or 1.0--7.9$\times10^{30}$\ergs , again
assuming a power-law or a blackbody, respectively.

\begin{figure*}
\hspace{0.2cm}
\hbox{
\includegraphics[height=7cm,width=8cm]{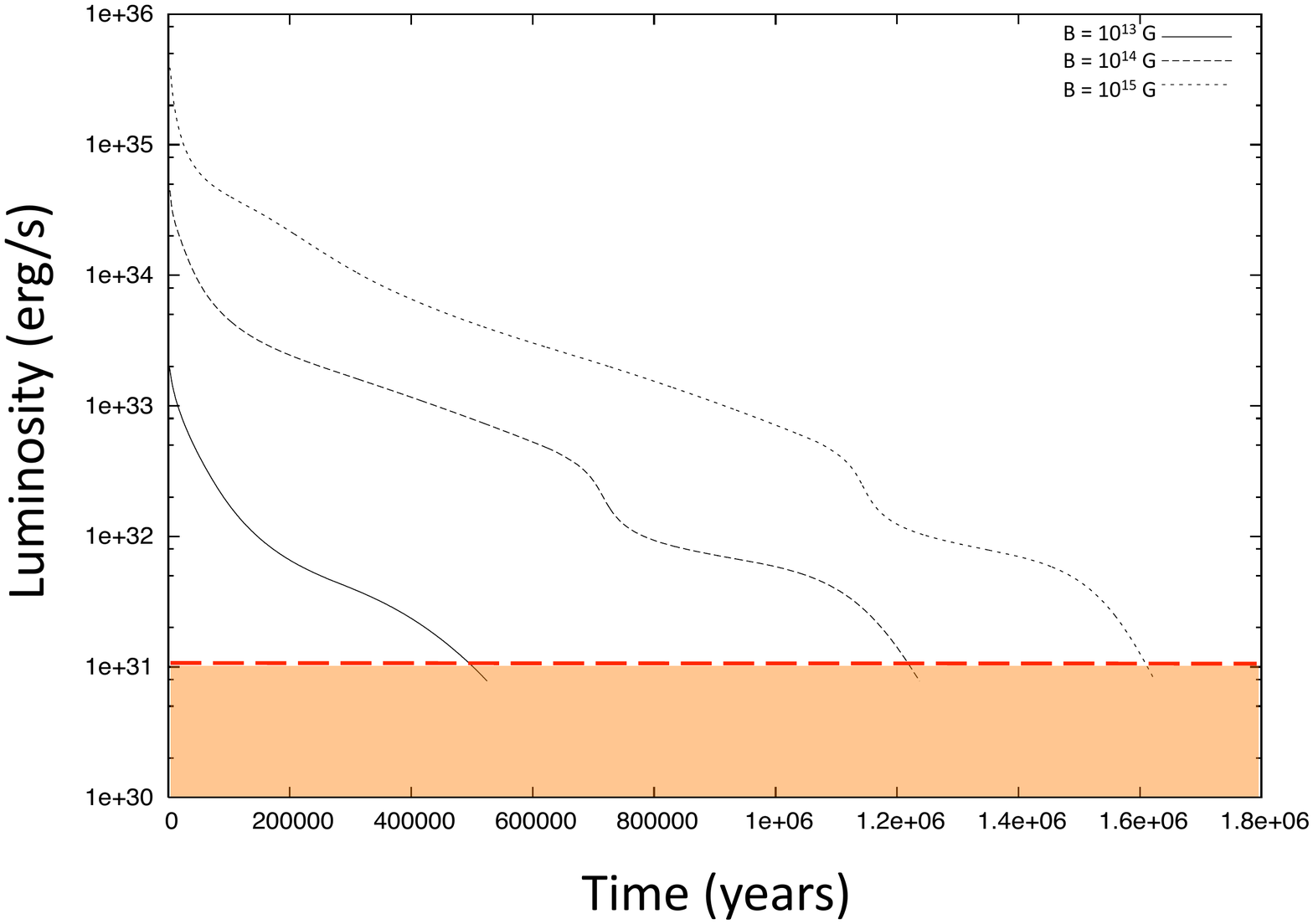}
\includegraphics[height=7cm,width=8cm]{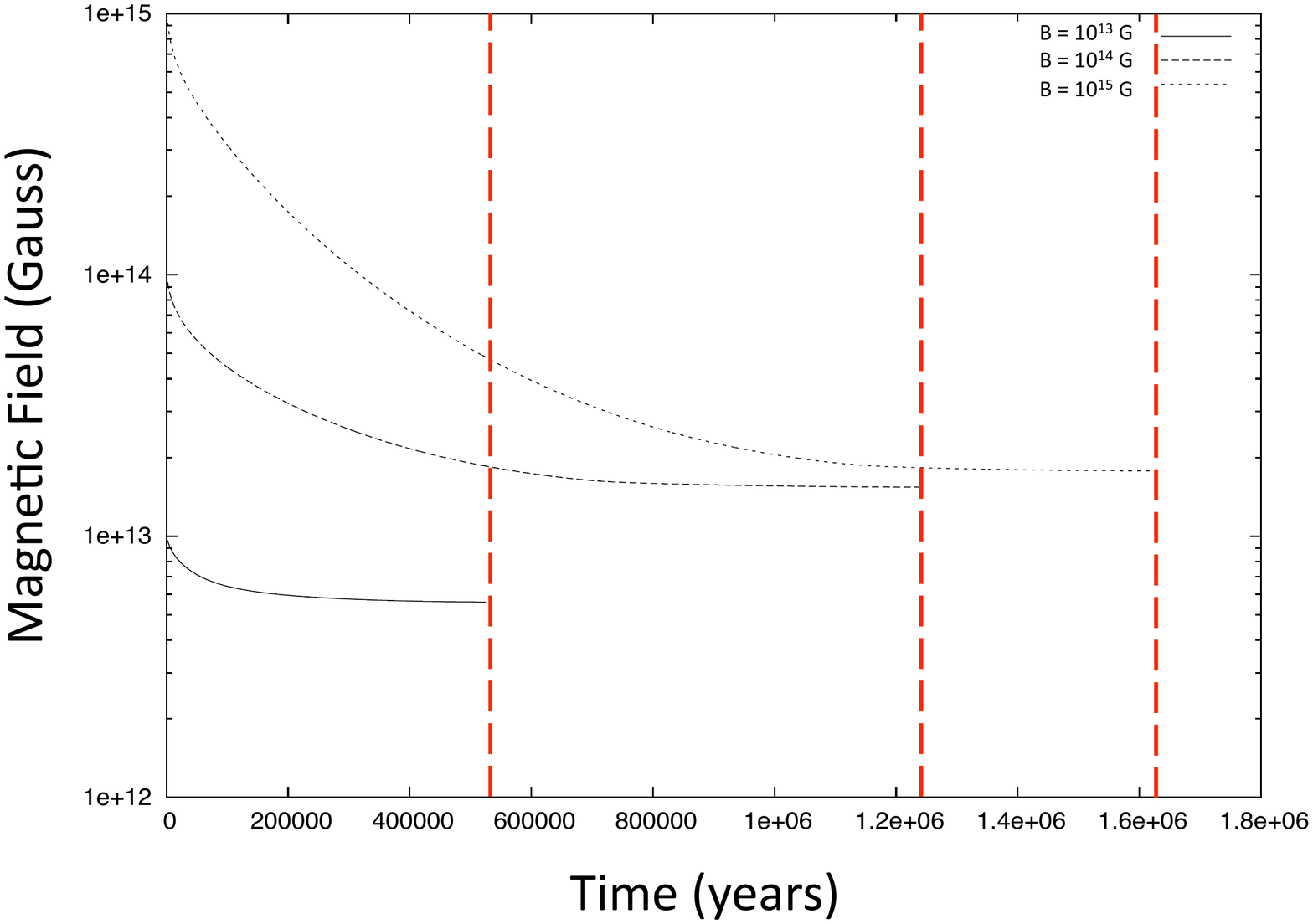}}
\caption{Luminosity decay (left panel) and magnetic field decay (right panel) of a 1.4\,$M_{\odot}$ neutron star with three initial "magnetar-like" magnetic fields (Pons, Miralles \& Geppert 2009). The orange region in the left panel corresponds to the luminosity limits we derived as a function of the distance (see text for details), with the maximum luminosity upper limit marked as a red dashed line (11.4$\times10^{30}$\ergs ). Vertical red-dashed lines in the right panel report on the age at which the corresponding cooling magnetar would reach this maximum luminosity upper limit.}
\label{pons}
\end{figure*}

\subsection{On the magnetar interpretation}

The discovery of fast optical bursts in \swt\, led a few authors to
claim the magnetar nature of this source (Castro-Tirado et al. 2008;
Stefanescu et al. 2008), based on the similarity of the optical
light-curve with the typical short X-ray bursts from magnetars.
However, neither a complete
understanding of the physical processes involved in these optical
bursts nor other magnetar-like features (slow spin period,
presence of magnetar-like X-ray bursts, etc.) are helping characterizing this source as a high magnetic field neutron stars. Furthermore, the GRB-like event emitted by this source is at variance with any other flaring activity detected thus far from magnetars (see Mereghetti 2008 for a review).

For an isolated neutron star, a luminosity limit of
$\sim10^{30}-10^{31}$ \ergs \, would necessarily imply a source older than
$5\times10^{5}$ years, for any cooling model or equation of state
(Lattimer \& Prakash 2001; Yakovlev \& Pethick 2004). This age limit
would make \swt\, two order of magnitudes older than the bulk of magnetars (Mereghetti 2008), in line only with the old low-B field \sgr\, (Rea et al. 2010). 
Furthermore, the former value can be
considered a lower limit on the age, since for the cooling models
for high magnetic field neutron stars (Pons, Miralles \& Geppert 2009;
Aguilera et al. 2008, 2009), and the heat released by the recent
outburst, these will go in the direction of predicting a larger age. In particular, for a typical cooling neutron star, the presence of a high magnetic field causes a much brighter
source at the same age.

Assuming a neutron star with 1.4~$M_{\odot}$, we have
used the cooling code of Pons, Miralles \& Geppert (2009) to simulate
the cooling decay of a magnetar with three different initial magnetic
field values (see Fig.\,\ref{pons} left panel). To reach our upper
limit on the luminosity of $L_{\rm x} \sim 10^{31}$\ergs \, for a
magnetar-like magnetic field of $B >10^{14}$\,G at birth, the source
should be now older than 120\, Myrs (see dashed line in Fig.\,\ref{pons} left panel), and having now a magnetic field of a few $\times10^{13}$, hence below the magnetar regime. On the other hand, the magnetic field and the strong internal helicity, supposed to produce short bursts and
outburst activity in magnetars,  should have been largely dissipated at these old times and low field (see Fig.\,\ref{pons} right panel). Taking at face value the luminosity
currently measured for typical quiescent magnetars, the luminosity we derive
is fainter than the faintest magnetar in quiescence
(\sgr : $\sim6\times10^{31}$\ergs ; Rea et al. 2010). The possibility of \swt\, being a case similar to the low-B field \sgr\, is intriguing, however the large GRB-like flare detected from the former would be hardly explainable within the scenario of an old magnetar releasing its last bit of internal magnetic energy through weak sporadic bursts, as for \sgr .

Furthermore, unless the source is in the Galactic halo at a
distance of $>10$\,kpc, it is also dimmer than the luminosity of any
X-ray Dim Isolated Neutron Star known to date ($\sim10^{31}$\ergs ;
Turolla 2009), making the possible association of this object with any
of those classes rather unlikely.

\begin{center}
\begin{figure*}
 \centerline{\includegraphics[height=9cm]{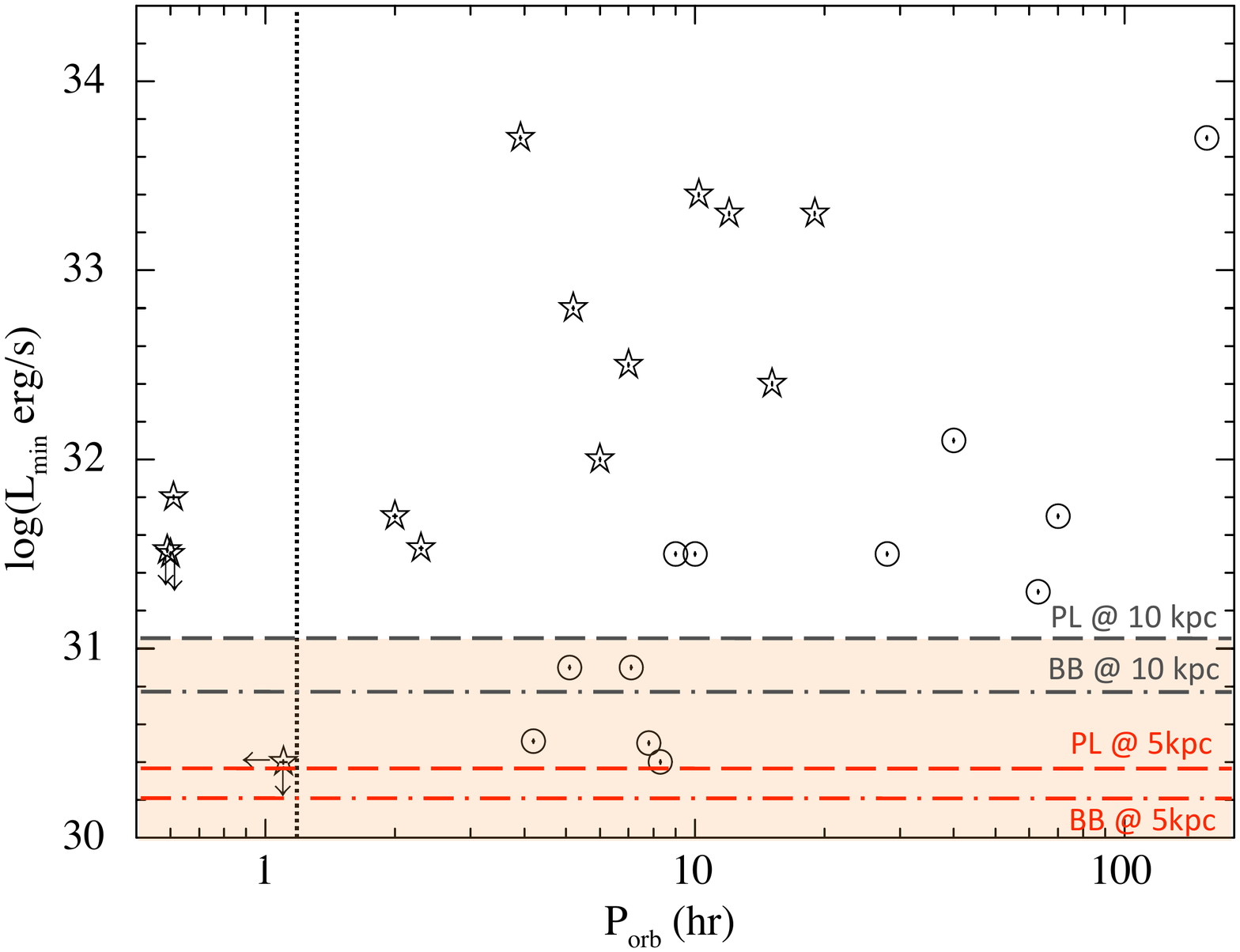}}
\caption{Quiescent 0.5--10\,keV luminosity versus P$_{\rm orb}$ for neutron star
  (stars) and black hole (circles) X-ray binaries (adapted from Lasota~2008;
  Garcia et al.~2001; Jonker et al.~2007; Degenaar et al. 2010; plus
  the addition of archival X-ray observations). Vertical arrows
  represent upper limits on the X-ray quiescent emission. Horizontal
  dashed and dot-dashed lines represent our 95\% upper limits on the
  X-ray quiescent luminosity of \swt\, for a power-law and a blackbody
  spectral models, respectively, and for a 5 and 10\,kpc distance (see text for details). 
  The orange shadowed region is the allowed luminosity space considering the whole 
  3.7--10\,kpc distance range, again for the two assumed spectral
  models. The vertical line at 1.2\,hr is the upper limit on the orbital
  period of \swt\, for a main-sequence companion star (see text more
  details).}
\end{figure*}
\end{center}

\subsection{An X-ray binary system}

A more plausible scenario is the X-ray binary nature of \swt. In the
binary case, optical and infrared observations during quiescence could
put a limit on the companion mass of $<$0.12~M$_{\odot}$ (with a
spectral type later than M5V), or being a semi-degenerate hydrogen
poor star (Castro-Tirado et al. 2008). The occurrence of the X-ray
outburst, implies that the low mass star orbiting the compact object
should be (close to) filling its Roche-Lobe. Assuming it fills its
Roche-Lobe, and it is a main sequence star, this gives a unique
relation between the orbital period of the system and the mass of
the companion star (see eq. 4.11 in Frank, King \& Raine~2002). 

Given the limits on the companion star, the system orbital
period is constrained to be shorter than 1.2\,hr if the star is on the
main sequence. Obviously, shorter periods are allowed also in the
ultra-compact binary case with an H-poor white dwarf. Another viable possibility is 
a hot brown dwarf companion star, in which case the orbital period of the system is constrained to be shorter than a few
hours (Bildsten \& Chakrabarty 2001). 


X-ray observations of low mass X-ray binaries during quiescence have
empirically shown that neutron star and black hole binaries, with
the same orbital periods, show different quiescent luminosities (see
e.g. Lasota 2008, and Fig.\,2). This observational evidence might have
a few interpretations. One possibility is that neutron stars' hot surface 
makes them always brighter during quiescence than a
black hole, where supposedly an event horizon is instead in place
(Narayan, Garcia \& McClintock 1997). Another possibility might
instead be the different accretion energy release mechanism, with
black holes releasing more energy through their radio jets rather than
in the X-ray band (Fender, Gallo \& Jonker 2003). In Fig.\,2 we plot
the quiescent X-ray luminosity of all binary neutron stars and
black holes for which this has been measured, and an orbital period or
a limit on it could be derived (see also Garcia et al.~2001; Kong et
al.~2002; Jonker et al.~2006; Lasota 2008). The orange region in Fig.\,2 is
the quiescent luminosity space limit we derived for \swt\, considering the
3.7--10\,kpc distance range and two different spectral models (see
above). We also plot the luminosity limits considering the most plausible distance of 5\,kpc, and for the larger distance of 10\,kpc (dashed and dot-dashed red and grey horizontal
lines, respectively).

In the rest of the discussion we attempt to distinguish between the neutron star and black hole hypotheses.

\subsubsection{A neutron star system}

The quiescent luminosity of a neutron star after $\sim$2 years from an
outburst is strongly dependent on the outburst history. The longer
the outburst activity the longer the cooling will take to reach the
pre-outburst luminosity level (Brown, Bildsten \& Rutledge 1998). In
our case, there are no previous outbursts or bursts recorded from
\swt\, in the past years, hence the heating dumped on the putative
neutron star surface is very little. This small heating
can explain the fast decrease in luminosity (by $\sim7$ orders of
magnitudes) from the GRB\,070610 event till August 2009 when our deep
X-ray upper limits are derived.

It is evident from Fig.\,2 that if at $\sim$5\,kpc, \swt\, would be an ultra-compact binary system, being too faint in quiescence for a neutron star accreting from a main-sequence star or a brown dwarf. In this scenario this system might be similar to the ultra-compact binary H1905+00 (Jonker et al. 2006, 2007; namely the upper limit reported on the bottom-left of the Fig.\,2). On the other hand, if the putative neutron star is instead at about 10\,kpc, then it might still be in orbit with a main sequence star or a brown dwarf, and  the binary should have an orbital period shorter than a few hours (Frank, King \& Raine~2002; Bildsten \& Chakrabarty 2001). 

That said, it is important to note that neutron star accreting systems
have not been seen showing large optical flares beside the optical
counterpart to Type I or II bursts which are orders of magnitude
fainter and with longer timescales than those observed in \swt\,
(Kasliwal et al. 2008; Stefanescu et al. 2008; Castro-Tirado et
al. 2008).

\subsubsection{A black hole system}

Large X-ray and optical flares, on several timescales, are an
ubiquitous characteristic of black hole binaries. In particular,
transient low-mass X-ray binaries hosting a black hole candidate undergo very dramatic X-ray and optical outbursts, and have long
periods (even decades) of quiescence. However, sub-second timescale
optical bursts such as in \swt\, (Stefanescu et al. 2008) were never
observed before in any black hole binary (nor any other astronomical
source either). Optical flares on several timescales down to minutes
were reported i.e. for A0620--00 (Hynes et al. 2003a), XTE\,J1118+480
(Hynes et al. 2003b), GRS 1124--684 and Cen X-4 (Shahbaz et
al. 2010). A somewhat similar case might be GX\,339-4, the typical
black hole candidate, which showed fast optical variability shorter than a
second during a recent outburst (Gandhi et al. 2010). Similarities can
be found also with the X-ray nova V4641 Sgr (as suggested by Kasliwal et
al. 2008), although in this case the relatively massive
companion star can introduce a somehow different physical process than in \swt .

Apart from the peculiar optical behavior, all the other observational characteristics observed from \swt\, are
in line with what already observed from black hole binaries: energetic
X-ray flares, fast decay into quiescence, optical QPOs at 0.16\,Hz, and 
extremely faint X-ray quiescence luminosities (see Fig.\,2).

Very interesting is the possibility of \swt\, being a black hole in an ultra-compact binary system, the first ever discovered. Although very speculative, it might be possible that the unique optical behavior of this source is indeed reflecting the first of such systems, the emission of which is still largely unknown.

\section{Conclusions}

We derived deep upper limits with \CXO\, on the X-ray quiescent emission of the optical bursting transient \swt . We showed that a magnetar scenario is very unlikely: the source is too faint in quiescence for any realistic scenario of magnetar cooling. We suggest that \swt\, is most likely an X-ray binary, hosting a black hole or a neutron star with an orbital period faster than a few hours, possibly in an ultra-compact system. High-time resolution optical observations of X-ray binaries during outburst might reveal energetic optical flares, a peculiarity that \swt\, does not share yet with any other source.

\acknowledgements
NR is supported by a Ram\'on~y~Cajal fellowship through Consejo Superior de Investigaciones Cientf\'icas, by grants AYA2009-07391 and SGR2009-811, and thanks P. Casella for useful discussions on the optical variability in black hole binaries. PGJ and GN acknowledges support from a VIDI grant from the Netherlands Organization for Scientific Research.

\end{document}